\documentclass[aps,twocolumn,prl]{revtex4}
\usepackage{latexsym}
\usepackage{amsmath}
\usepackage{mathrsfs}
\usepackage{amsthm}
\usepackage{amssymb}
\usepackage{fancybox}
\usepackage{bm}
\usepackage{fancyhdr}
\usepackage{subfigure}
\usepackage{graphicx}
\usepackage{color}

\begin{document}

\title{PT-Symmetric Oligomers: Analytical Solutions, Linear Stability and Nonlinear
Dynamics}
\author{K. Li}
\affiliation{
Department of Mathematics and Statistics, University of Massachusetts, Amherst MA 01003-4515, USA}
\author{P.G. Kevrekidis}
\affiliation{
Department of Mathematics and Statistics, University of Massachusetts, Amherst MA 01003-4515, USA}

\begin{abstract}
In the present work we focus on the case of (few-site) configurations 
respecting the
PT-symmetry. We examine the case of such ``oligomers'' with not only 2-sites,
as in earlier works, but also the cases of 3- and 4-sites. While
in the former case of recent experimental interest, the picture
of existing stationary solutions and their stability is fairly 
straightforward, the
latter cases reveal a considerable additional complexity of solutions,
including ones that exist past the linear PT-breaking point
in the case of the trimer, and more complex, even asymmetric solutions
in the case of the quadrimer with nontrivial spectral and 
dynamical properties.
Both the linear stability and the nonlinear dynamical properties
of the obtained solutions are discussed.
\end{abstract}

\maketitle

\section{Introduction}

Over the past decade, the examination of Hamiltonian nonlinear dynamical
lattices, as well as that of continuum systems with periodic
potentials has been a subject of intense investigation \cite{reviews}.
The motivation for such studies stems from a variety of physical
settings including, among others, the themes of optical beam
dynamics in coupled waveguide arrays or optically induced photonic
lattices in photorefractive crystals \cite{reviews1},
the temporal evolution of Bose-Einstein
condensates (BECs) in optical lattices \cite{reviews2}, or the
DNA double strand denaturation in biophysics \cite{reviews3}.
One of the common focal points among all of these areas has been
the intense study of the existence, stability and dynamical properties
of their nonlinear (often localized in the form of solitary waves)
solutions which are of principal interest and experimental observability
within various applications; see \cite{pgk_rev} for a relevant
recent review.

On the other hand, as the understanding of the conservative
aspects of such systems comes to a point of maturation, a number
of interesting variants thereof arise. A canonical one concerns
the examination of effects of damping and driving that
not only yield novel theoretical solutions (see as an example
\cite{maniadis}), but also are inherently relevant to applications
(again, see for a recent example \cite{lars}). A more exotic
variant which, however, in the past couple of years has gained
considerable momentum especially due to the recent experiments
of \cite{kip} is that of PT-symmetric dynamical lattices.
This theme follows the pioneering realization of Bender and
coworkers \cite{bend} that non-Hermitian Hamiltonians can
still yield real spectra, provided that they respect the Parity (P)
and Time-reversal (T) symmetries. Practically, in the presence
of a (generally complex) potential the relevant transformations
imply that the potential satisfies the condition $V(x)=V^{\star}(-x)$.
In nonlinear optics, the interest in such applications was initiated
by the key contributions of Christodoulides and co-workers \cite{christo1}
which considered solitary waves as well as linear (Floquet-Bloch)
eigenmodes in periodic potentials satisfying the above condition,
also including the effects of Kerr nonlinearity and observing how
the properties of such waves were modified by genuinely complex,
yet PT-symmetric potentials.

More recently, motivated by the experimental possibilities and
the relevant realization of a PT-``coupler'' in \cite{kip},
there has been an interest in merging the experience of the
above two areas, namely the consideration of PT-symmetric
settings but for genuinely discrete media. In that vein,
the experimentally-probed two-site system has been considered
in the work of \cite{kot1}, where it was shown that it can operate
as a unidirectional optical valve, as well as in the study of
\cite{sukh1}, where the role of nonlinearity in allowing
(if sufficiently weak) or suppressing (if sufficiently strong)
time reversals of exchanges of optical power between the sites.
Another recent example consisted of the generalization of
\cite{kot2} where a lattice of coupled gain-loss dimers
was considered. This theme has also been considered in the
BEC literature and in the context of the so-called leaky
Bose-Hubbard dimers (allowing e.g. the tunneling escape of atoms
from one of the wells of a double-well potential). 
There, a variant of the model considered
below has been self-consistently derived in the mean-field
approximation \cite{grae1} and the correspondence of its
classical with the full quantum behavior has been explored \cite{grae2}.

Our aim in the present work is to revisit the examination of
the PT-symmetric coupler and to give a simple and complete
characterization of the existence and stability properties
of its stationary solutions. It should be noted that this aspect has been
partly addressed in both \cite{kot1} and \cite{sukh1}. Nevertheless,
we aim to give a characterization thereof
as a preamble towards the more complex (and thus, arguably,
more interesting)
generalization to what we call ``PT-symmetric oligomers'', namely
the consideration of a PT-symmetric trimer and that of a PT-symmetric
quadrimer. Our aim here is to explore how the complexity of the problem
expands as more sites are added, in order to offer a glimpse how
such oligomers gradually give way to the elaborate phenomenology of
a PT-symmetric lattice. We illustrate, for example, how it is possible
in the case of a trimer to identify stationary solutions which exist past the
limit of linear PT-symmetry breaking (something which is not possible
in the dimer case). We then proceed to illustrate how the phenomenology
of the quadrimer is even richer and more complex, featuring among
others asymmetric solutions with a reduced symmetry spectrum
differently than is the case for both the dimer and trimer.

Our presentation will be structured as follows. In section II,
we consider the fundamental (and previously considered)
dimer case. We use this as a benchmark for the
presentation of our methods and results. We then turn to the
more complex trimer case in section III and conclude our results
with section IV on the quadrimer. Finally, section V summarizes
our findings and presents some interesting questions for further
study.

\section{Dimer}

We start our considerations from the so-called PT-symmetric
coupler or dimer (as we will call it hereafter). In this
case, the dynamical equations are of the form:
\begin{eqnarray}
i\dot{u}_1&=&-ku_2-|u_1|^2u_1-i\gamma u_1
\nonumber
\\
i\dot{u}_2&=&-ku_1-|u_2|^2u_2+i\gamma u_2.
\label{dim1}
\end{eqnarray}
The model of Eq.~(\ref{dim1}) considers the linear PT-symmetric
dimer experimentally examined in \cite{kip}, as augmented by
the Kerr nonlinearity relevant e.g. to optical waveguides; see also
\cite{kot1,sukh1}. The overdot denotes the derivative with respect
to the evolution variable which in optical applications is the propagation
distance. In what follows, we will denote this variable by $t$ (to indicate
its evolutionary nature).
We seek stationary solutions of the form $u_1=\exp(i E t) a$ and
$u_2=\exp(i E t) b$. Then the stationary equations arise:
\begin{eqnarray}
E a&=&k b+|a|^2a+i\gamma a
\nonumber
\\
E b&=&k a+|b|^2b-i\gamma b.
\label{dim2}
\end{eqnarray}

Using a generic polar representation of the two ``sites''
$a=Ae^{i\phi_a}, b=Be^{i\phi_b}$, we are led to the following
algebraic conditions for the two existing branches of solutions
(notice the $\pm$ sign distinguishing between them):
\begin{eqnarray}
A^2=B^2=E \pm \sqrt{k^2-\gamma^2}
\label{dimer3}
\\
\sin(\phi_b-\phi_a)=-\frac{\gamma}{k}
\label{dimer4}
\end{eqnarray}
The fundamental difference of such solutions from their
standard Hamiltonian ($\gamma=0$) counterpart is that the
latter were lacking the ``flux condition'' of Eq.~(\ref{dimer4}).
This dictated a selection of the phases so that no phase
current would arise between the sites. On the contrary,
in PT-symmetric settings, the phase flux is nontrivial
and must, in fact, be consonant with the gain-loss pattern
of the coupler.

\begin{figure}[htp]
\scalebox{0.27}{\includegraphics{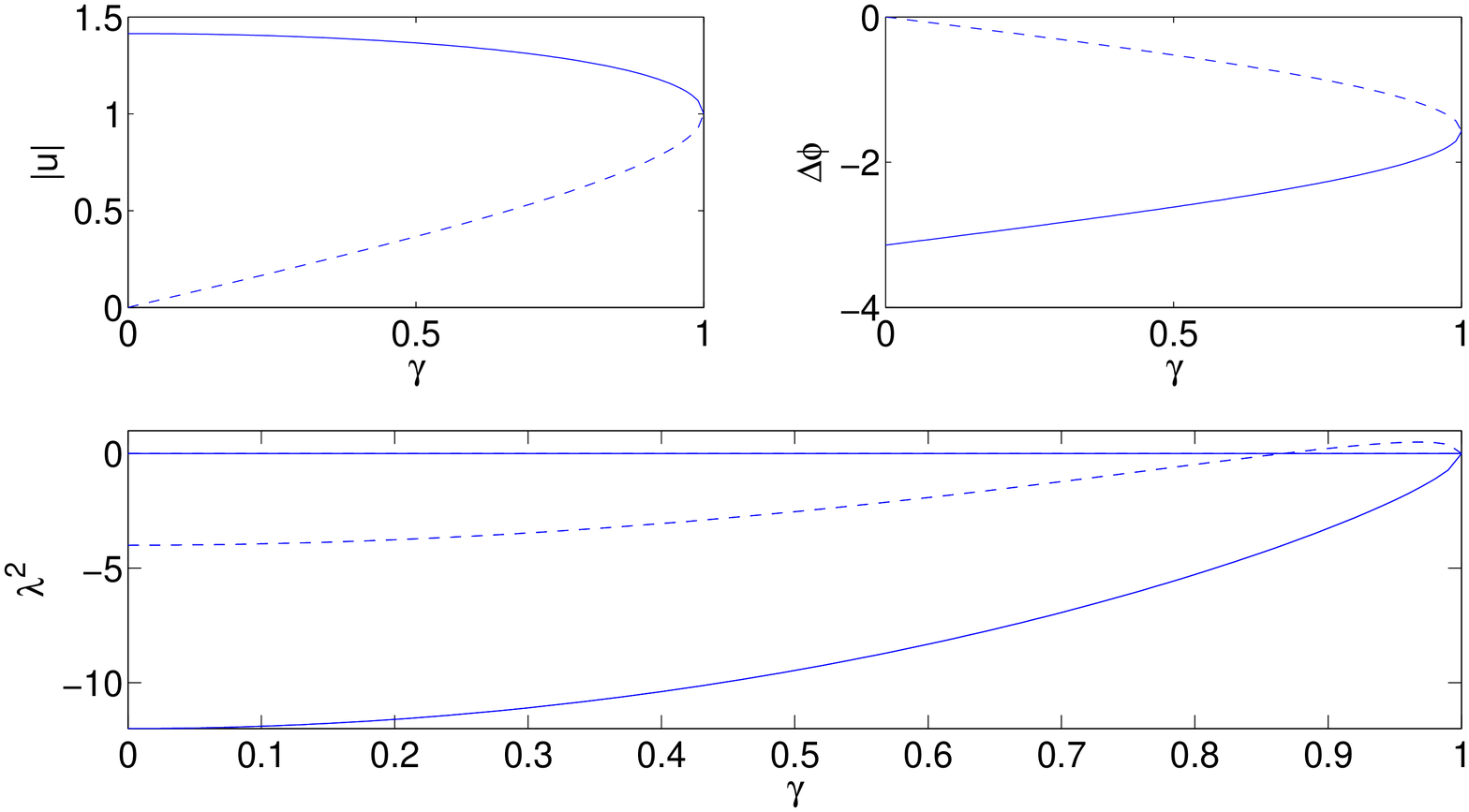}}
\caption{The two branches of solutions for the dimer problem are shown
for parameter values $k=E=1$. The top left illustrates the amplitude
of the sites, the top right their relative phase and the bottom panel shows
the (nontrivial) squared eigenvalue of the two branches. The solid line
corresponds to the always stable branch $u^{(2)}$, while the
dashed branch corresponds to $u^{(1)}$, which acquires
a real eigenvalue pair above a certain $\gamma=\sqrt{k^2-E^2/4}$.}
\label{dimer}
\end{figure}

Fig.~\ref{dimer} shows the profile of the two branches.
The first branch $u^{(1)}$ corresponding to the $(-)$ sign in Eq.~(\ref{dimer3}) is stable when
$\gamma^2 \le k^2-E^2/4$, whereas the second branch $u^{(2)}$ is always stable.
The linearization around these branches can be performed explicitly
yielding the nonzero eigenvalue pairs
$\pm 2 i \sqrt{2(k^2-\gamma^2)-E\sqrt{k^2-\gamma^2}}$
for the first and $\pm 2 i \sqrt{2(k^2-\gamma^2)+E\sqrt{k^2-\gamma^2}}$ for the
second (notice that the latter can never become real).

\begin{figure}[htp]
\scalebox{0.28}{\includegraphics{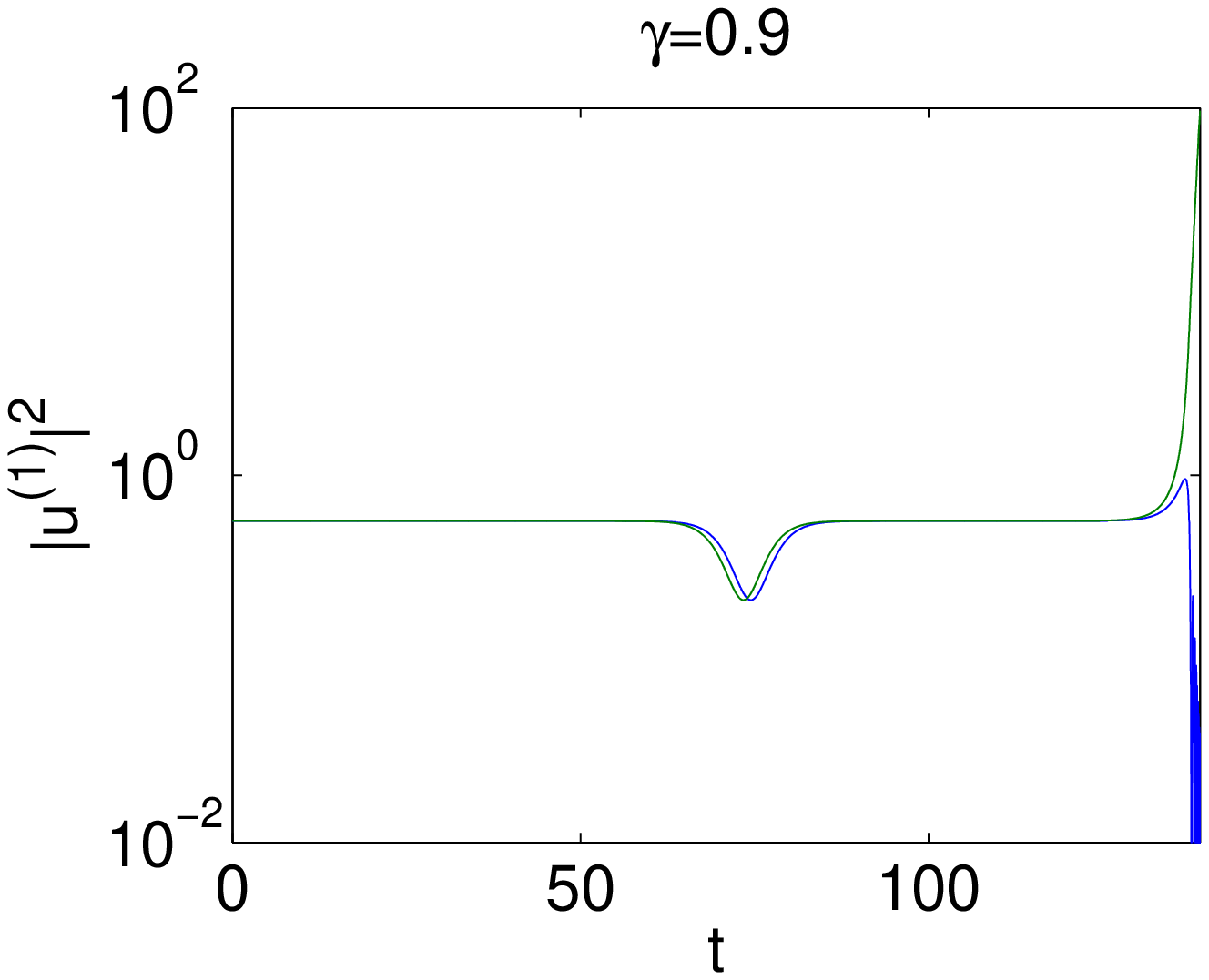}}
\scalebox{0.28}{\includegraphics{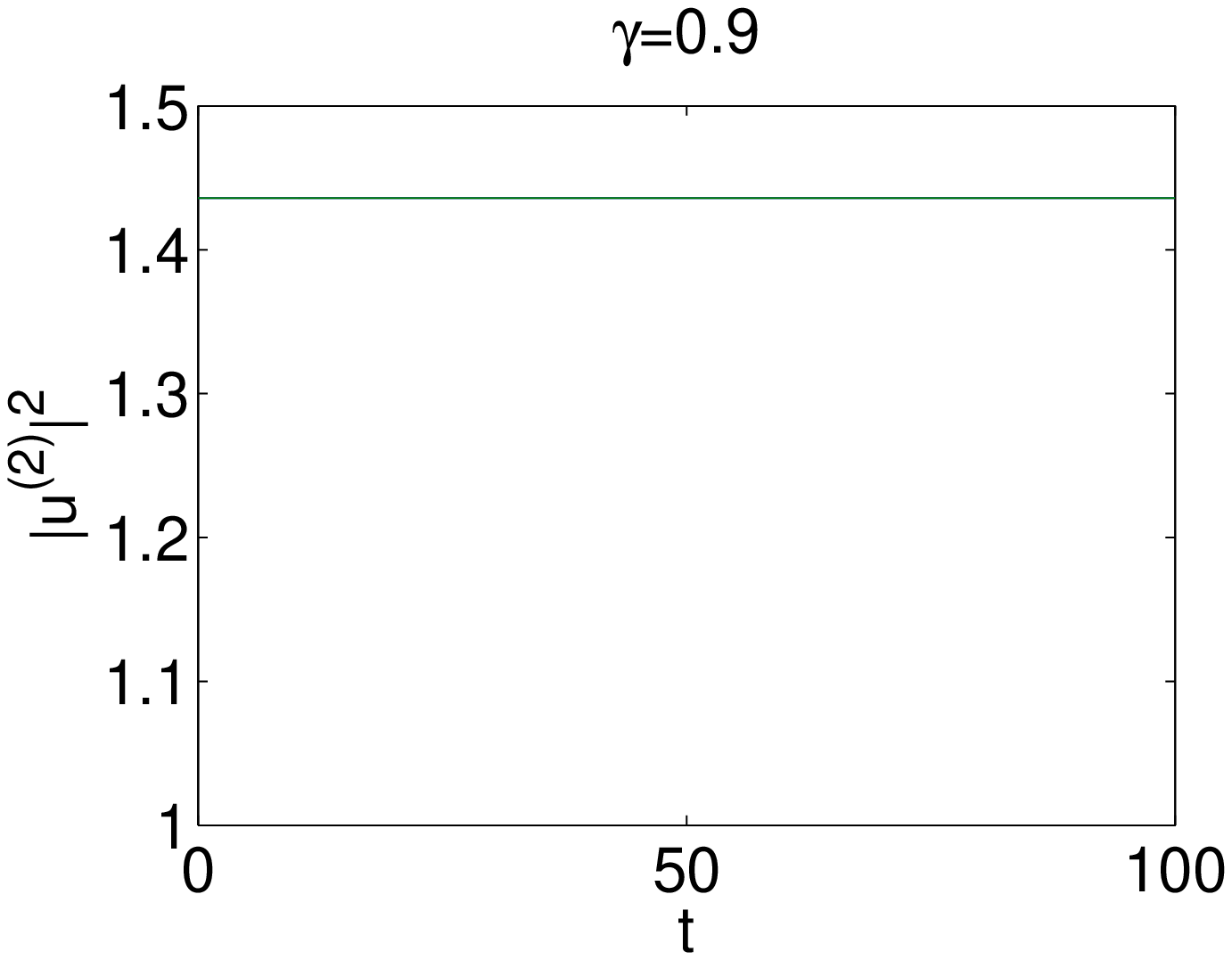}}
\caption{Dynamical evolution of initial data belonging to the
two branches of stationary solutions of a dimer in the case of
$\gamma=0.9, \ E=k=1$, which is past the critical point for
the instability of the first branch (left panel), while the
second branch of the right panel is still dynamically stable.
Notice that the left panel is plotted in semilog.}
\label{dimer_r1}
\end{figure}

\begin{figure}[htp]
\scalebox{0.28}{\includegraphics{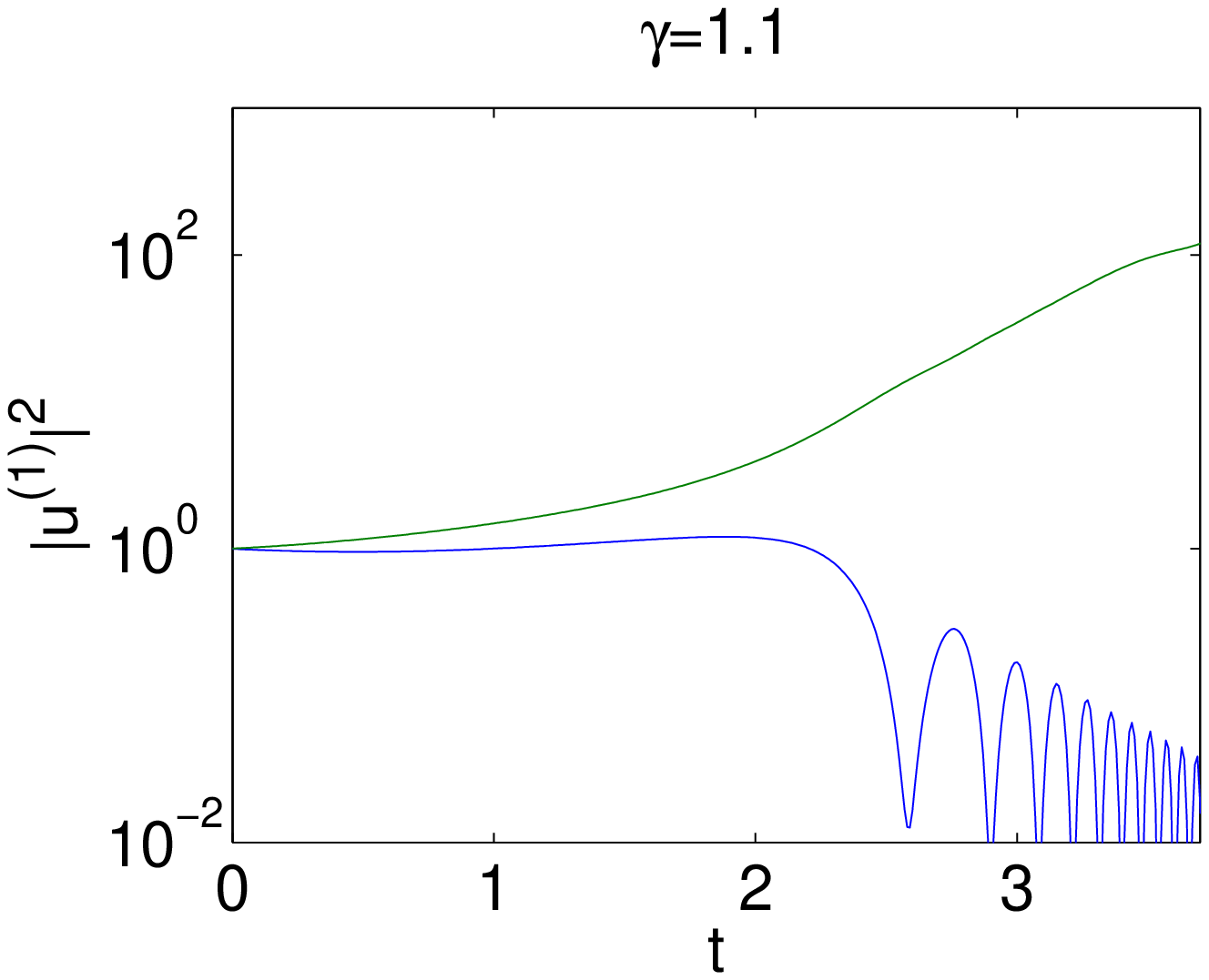}}
\scalebox{0.28}{\includegraphics{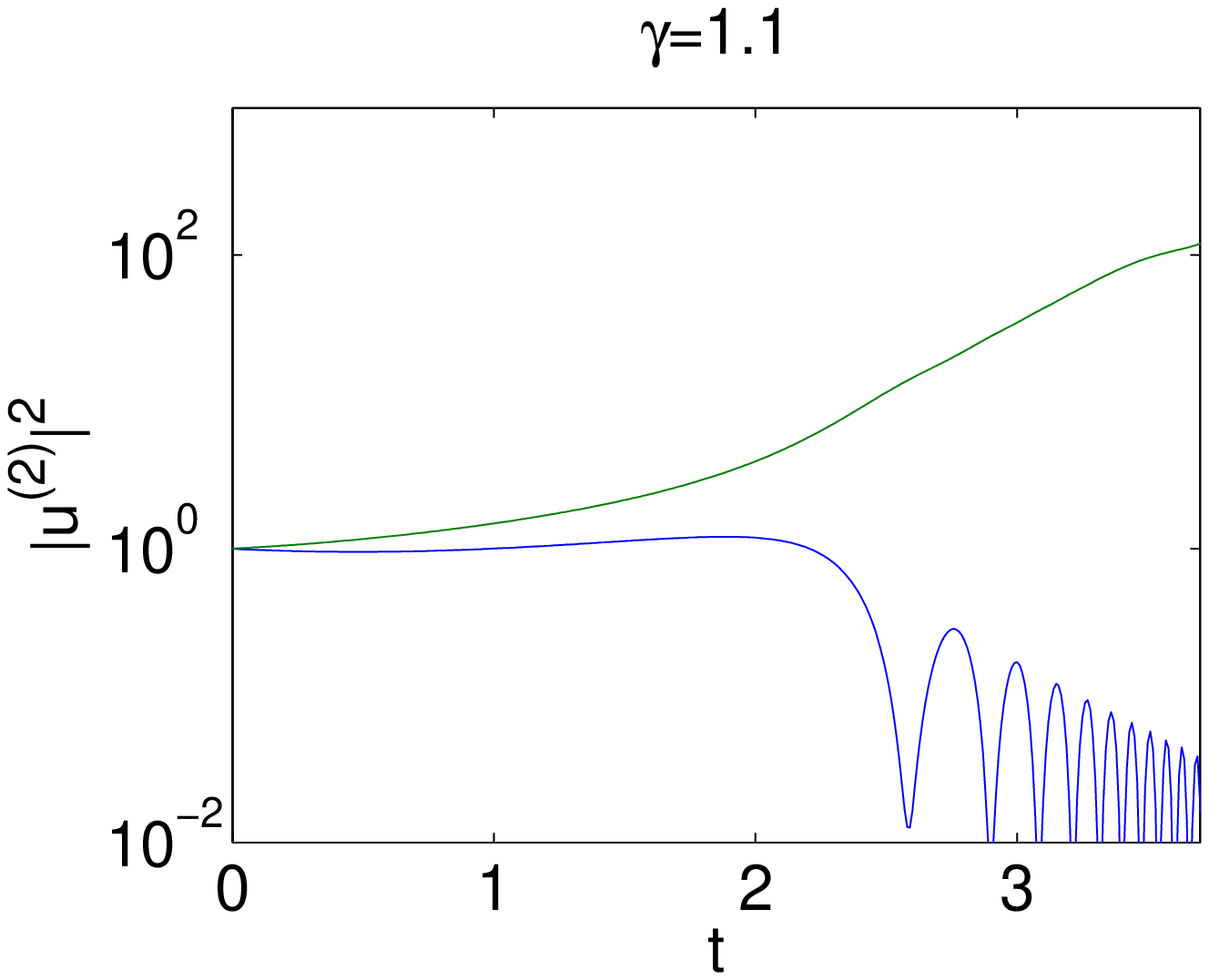}}
\caption{Similar initialization of the dimer based on the two
branches of stationary solutions (for $\gamma=1$) but now for
the case of $\gamma=1.1$ ($E=k=1$). The asymmetric evolution of the
coupler past the linear PT-symmetric threshold can be clearly discerned.}
\label{dimer_r2}
\end{figure}

It is relevant to note here that the two branches ``die'' in a
saddle-center bifurcation at $\gamma=k$, as shown in the figure.
Importantly, this coincides with the linear limit of the
PT-symmetry breaking since the linear eigenvalues of the
problem are $\lambda=\pm \sqrt{k^2-\gamma^2}$. Hence, the nonlinear
solutions terminate where the linear problem eigenfunctions
yield an imaginary pair, predisposing us for
an asymmetric evolution past this critical point (for all
initial data). The dynamical evolution of the dimer is shown
first for a case of $\gamma<k$
(in which $u^{(1)}$ is unstable, while $u^{(2)}$ is stable)
in Fig.~\ref{dimer_r1}.
The evolution of the instability of $u^{(1)}$
leads to an asymmetric distribution of the power in the coupler,
despite the fact that parametrically we are below the
linear critical point (for the PT-symmetry breaking). 
Notice that in all the cases, also below, where a stationary
solution exists for the parameter values for which it is initialized,
dynamical instabilities arise only through the amplification of
roundoff errors i.e., a numerically exact solution up to 
$10^{-8}$ is typically used as an initial condition in the system.
Naturally,
beyond $\gamma=k$, as shown in Fig.~\ref{dimer_r2}, all
initial data yield such an asymmetric evolution. 

\section{Trimer}

We now turn to the case of the trimer where the dynamical equations are
\begin{eqnarray}
i\dot{u}_1&=&-ku_2-|u_1|^2u_1-i\gamma u_1
\nonumber
\\
i\dot{u}_2&=&-k(u_1+u_3)-|u_2|^2u_2
\nonumber
\\
i\dot{u}_3&=&-ku_2-|u_3|^2u_3+i\gamma u_3
\label{trimer1}
\end{eqnarray}
Seeking once again stationary solutions leads to the algebraic equations
\begin{eqnarray}
E a&=&k b+|a|^2a+i\gamma a
\nonumber
\\
E b&=&k (a+c)+|b|^2b
\nonumber
\\
E c&=&k b+|c|^2c-i\gamma c
\label{trimer2}
\end{eqnarray}

In this case too, it is helpful to use the polar representation
for the three-sites in the form
$a=Ae^{i\phi_a}, b=Be^{i\phi_b}, c=Ce^{i\phi_c}$, which, in turn,
leads to the algebraic equations of the form:
\begin{eqnarray}
A=C
\label{trimer3}
\\
B^4-EB^2+2EA^2-2A^4=0
\label{trimer4}
\\
\sin(\phi_b-\phi_a)=-\sin(\phi_b-\phi_c)=-\frac{\gamma A}{k B}
\label{trimer5}
\\
\cos(\phi_a-\phi_b)=\cos(\phi_b-\phi_c)=\frac{EA-A^3}{k B}
\label{trimer6}
\end{eqnarray}
Notice how the presence of the gain-loss spatial profile along
the 3-sites induces a spatial phase distribution and enforces
the condition of a symmetric amplitude profile with the first
and third site sharing the same amplitude.
This phase distribution would be trivial (relative phases of
$0$ or $\pi$) in the $\gamma=0$ case.

\begin{figure}[htp]
\scalebox{0.28}{\includegraphics{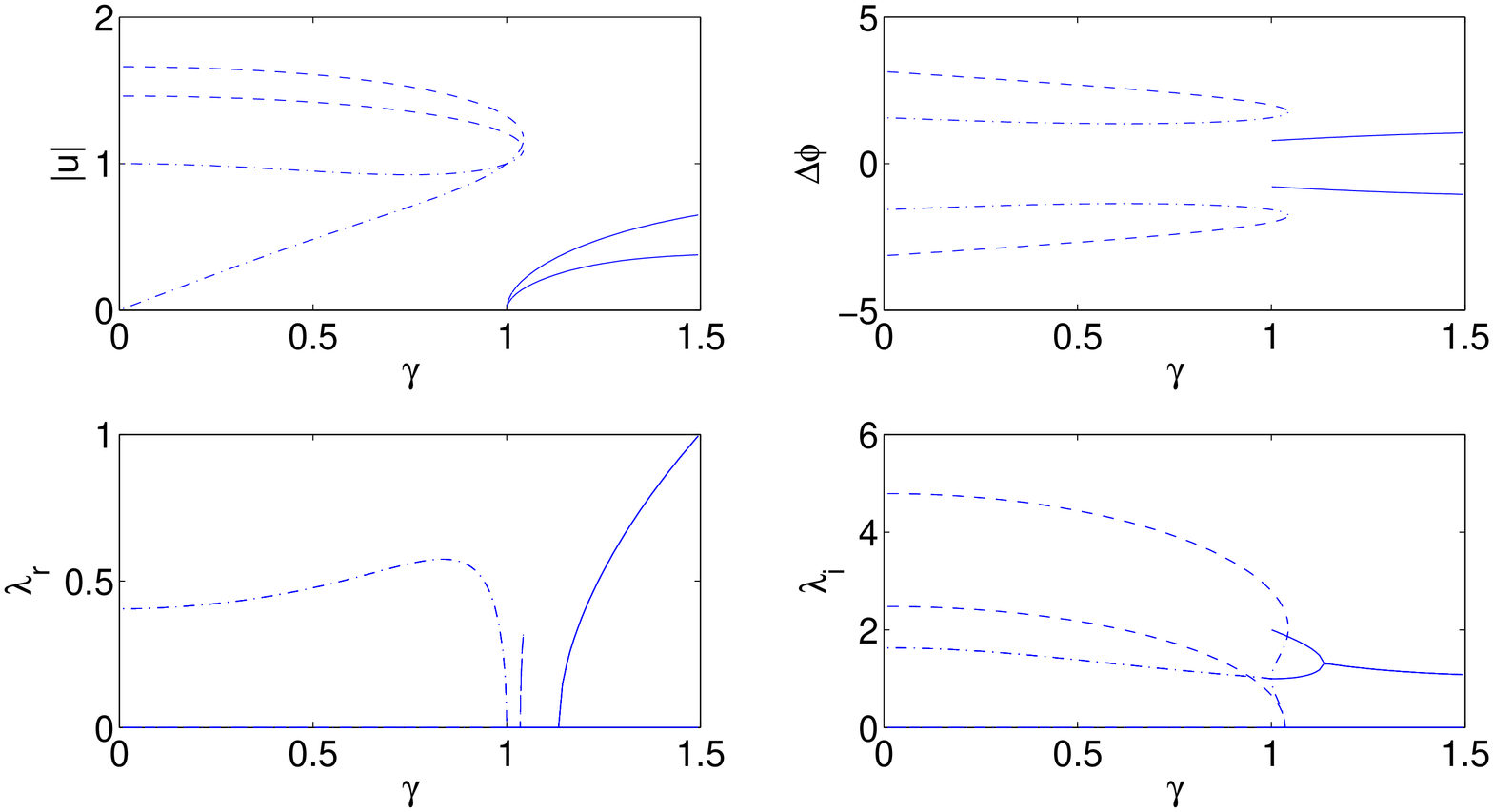}}
\caption{Existence and stability of solutions for the case of the
trimer analogously to Fig.~\ref{dimer} for the dimer case. The
only other difference is the illustration of both real and imaginary
parts for the eigenvalues in the bottom panel.
We have set (without loss of generality)
$\phi_b$ normalized to $0$. Different line styles are used to
distinguish different branches and their corresponding stabilities (see the
detailed explanation in the text).}
\label{trimer}
\end{figure}

A typical example of the branches that may arise in the case of the
trimer is shown in Fig.~\ref{trimer} for $E=k=1$. In this case, we find
three distinct branches in the considered interval of parameter values.
There are two branches which exist up to the critical point
$\gamma=1.043$. In this interval one of the two branches $u^{(1)}$
is chiefly unstable (denoted by dash-dotted line) except for
a small interval of $\gamma \in [1,1.035]$. The other one $u^{(2)}$
is chiefly stable (denoted by a dashed line)
except for $\gamma \in [1.035,1.043]$.
The eigenvalues of $u^{(1)}$ and $u^{(2)}$ in
$\gamma \in [1.035,1.043]$ are very close to each other but not
identical.
Notice that $u^{(1)}$ is unstable due to a complex
eigenvalue quartet whose eigenvalues collide on the imaginary
axis for $\gamma=1$ and split into two imaginary pairs one of which
becomes real for $\gamma>1.035$. Finally, these two branches
collide in a saddle-center bifurcation (for $\gamma=1.043$)
and disappear thereafter.


Interestingly, however, these are not the only branches that
arise in the trimer case. In particular, as can be seen
in Fig.~\ref{trimer}, there is a branch of solutions
bifurcating from zero (amplitude) for $\gamma > \sqrt{2k^2-E^2}$, denoted by
$u^{(3)}$, the solid line in Fig.~\ref{trimer}. In our case $E=k=1$,
this branch is only stable for $\gamma< 1.13$, at which
point two pairs of imaginary eigenvalues collide and lead
to a complex quartet which renders the branch unstable thereafter.
Yet, this branch of solutions has a remarkable trait.
In the case of the trimer, the underlying linear problem possesses
the following eigenvalues
$0$, $\pm \sqrt{2 k^2-\gamma^2}$. Hence, the critical point
for the existence of real eigenvalues of the linear problem
in the case of the PT-symmetric trimer is $\gamma=\sqrt{2} k$
(cf. with the $\gamma=k$ limit of the dimer). Nevertheless,
and contrary to what is the case for the dimer, the third
branch of solutions considered above persists beyond this
critical point (although it is unstable in that regime).

\begin{figure}[htp]
\scalebox{0.28}{\includegraphics{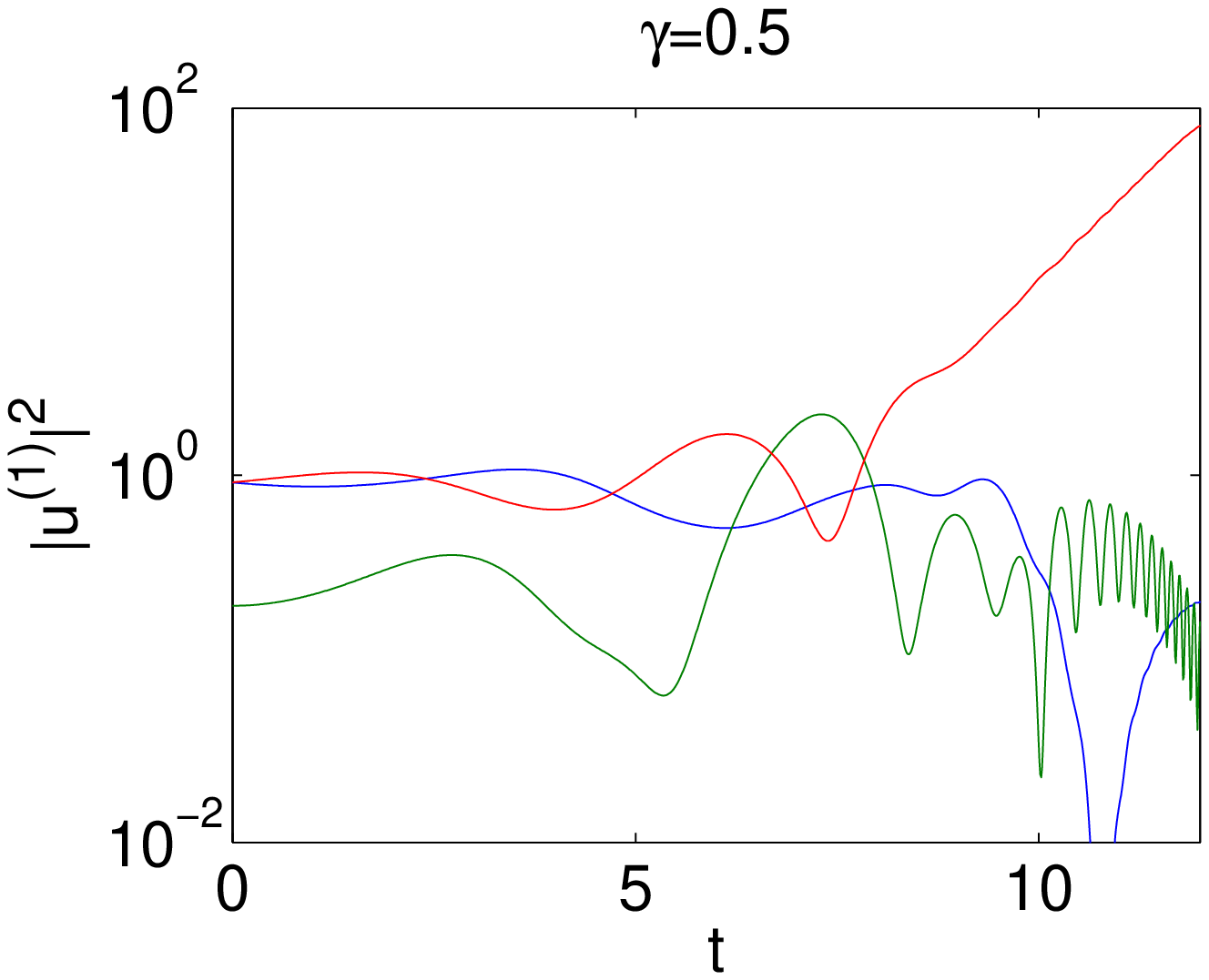}}
\scalebox{0.28}{\includegraphics{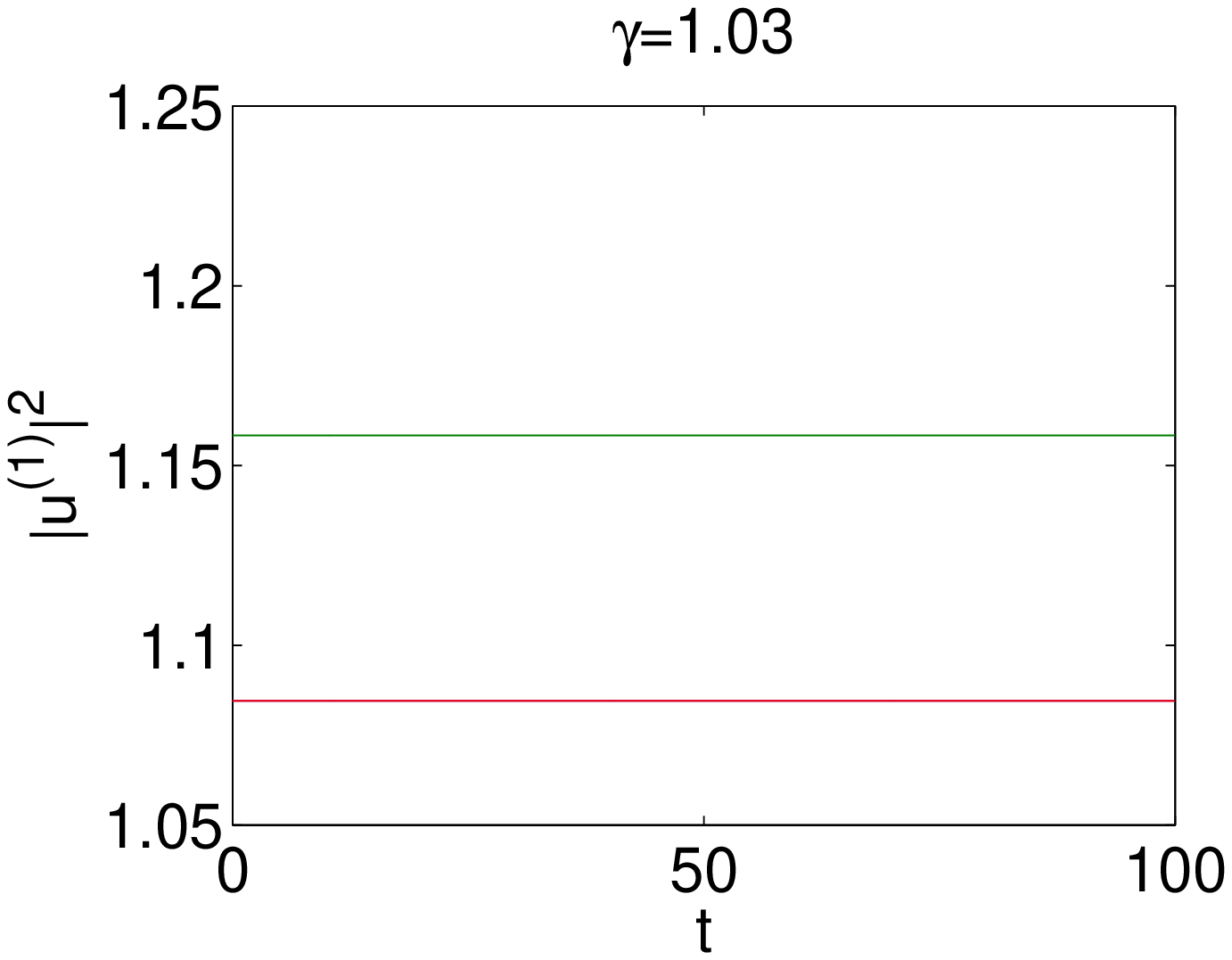}}
\scalebox{0.28}{\includegraphics{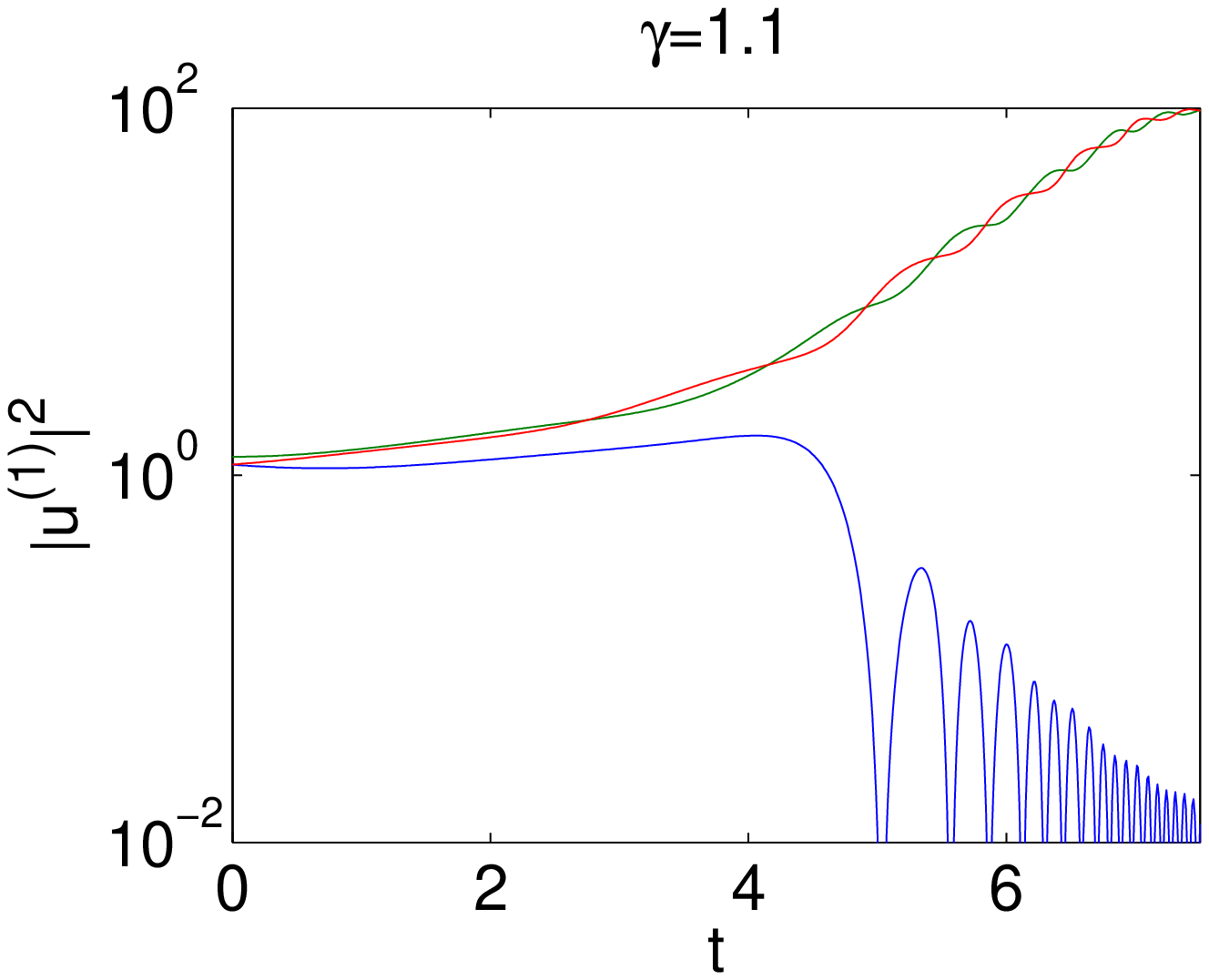}}
\caption{The dynamical evolution of the amplitudes of
the three sites of the stationary solution $u^{(1)}$ in the case of $E=k=1$
for $\gamma=0.5$ (left panel), $\gamma=1.03$ (right panel)
and $\gamma=1.1$ (bottom panel).
The bottom panel is initialized with the exact
stationary solution for $\gamma=1.04$ (since
for $\gamma=1.1$ the branch no longer exists as a stationary
solution).}
\label{trimer_u1}
\end{figure}

\begin{figure}[htp]
\scalebox{0.28}{\includegraphics{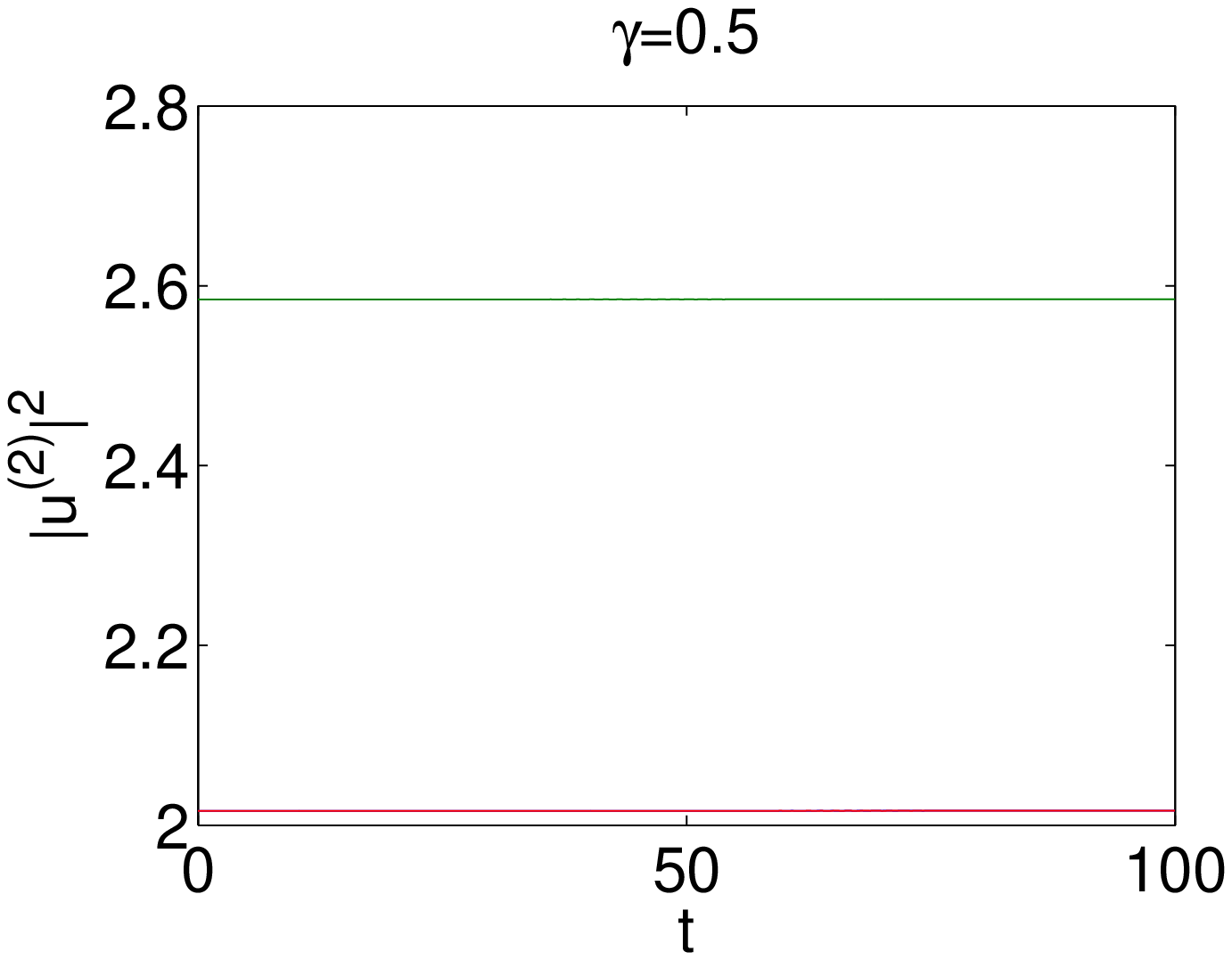}}
\scalebox{0.28}{\includegraphics{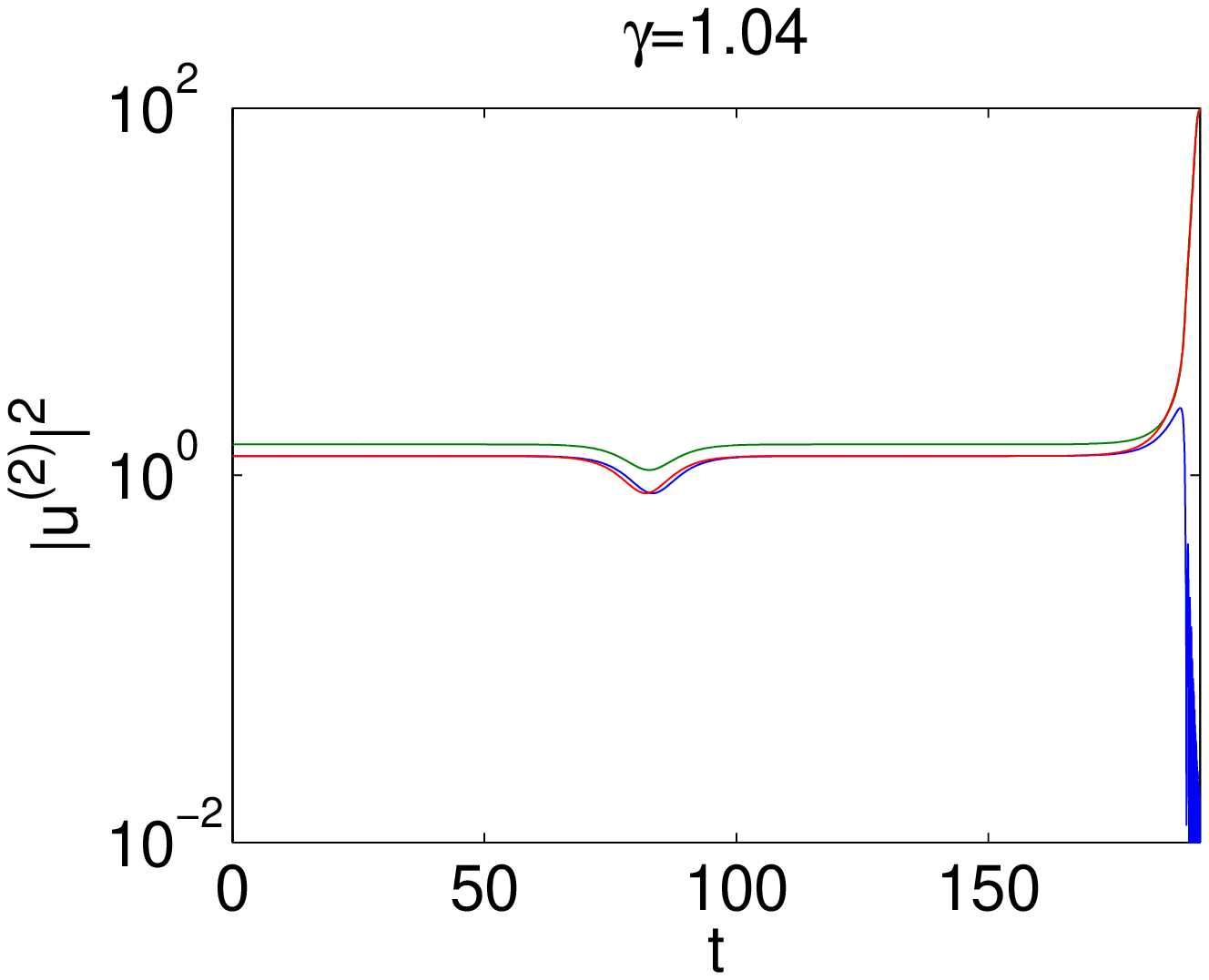}}
\scalebox{0.28}{\includegraphics{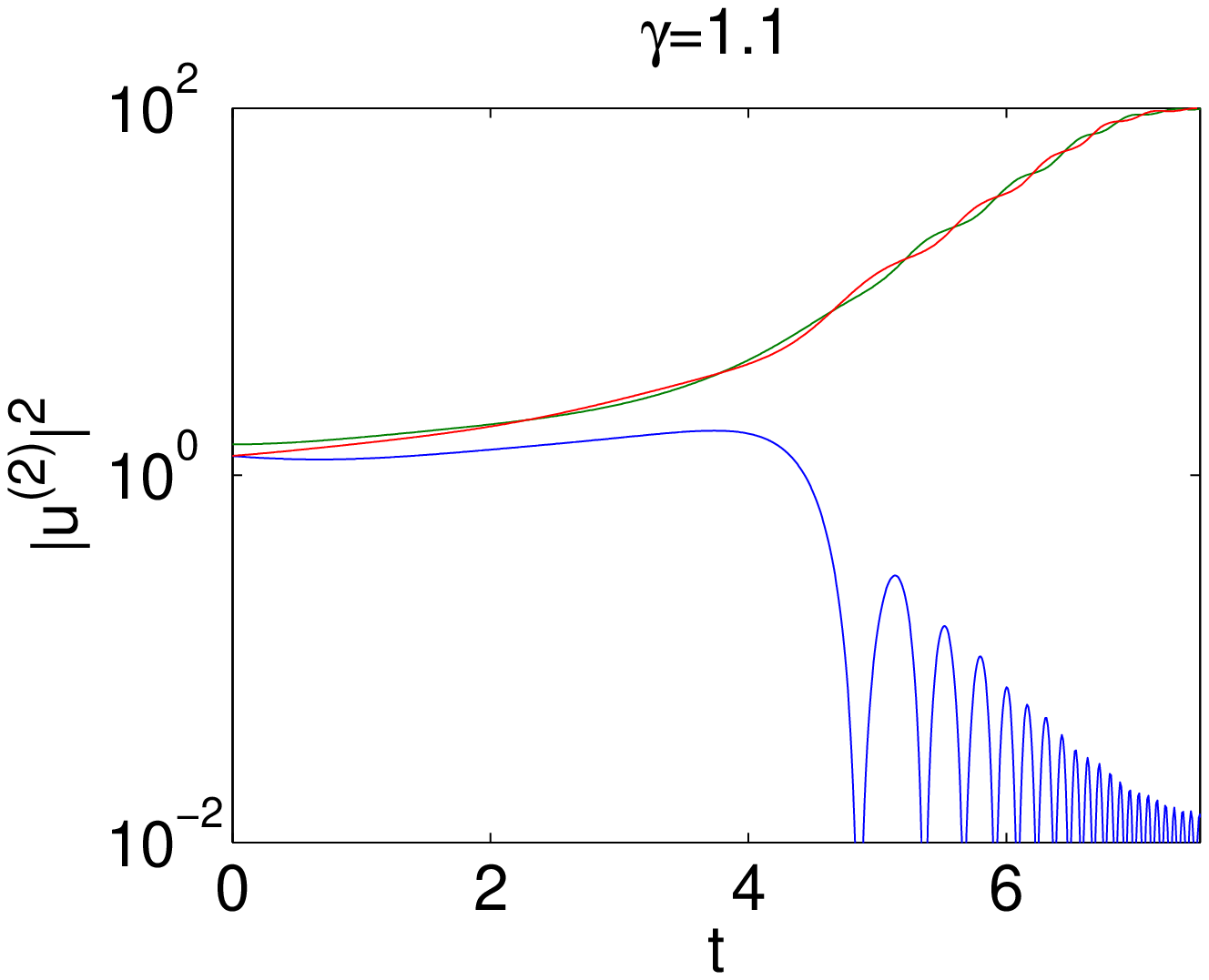}}
\caption{Similar to Fig. \ref{trimer_u1} but for the case of
$u^{(2)}$ and for $E=k=1$.
The bottom panel is again initialized for $\gamma=1.04$.}
\label{trimer_u2}
\end{figure}

\begin{figure}[htp]
\scalebox{0.28}{\includegraphics{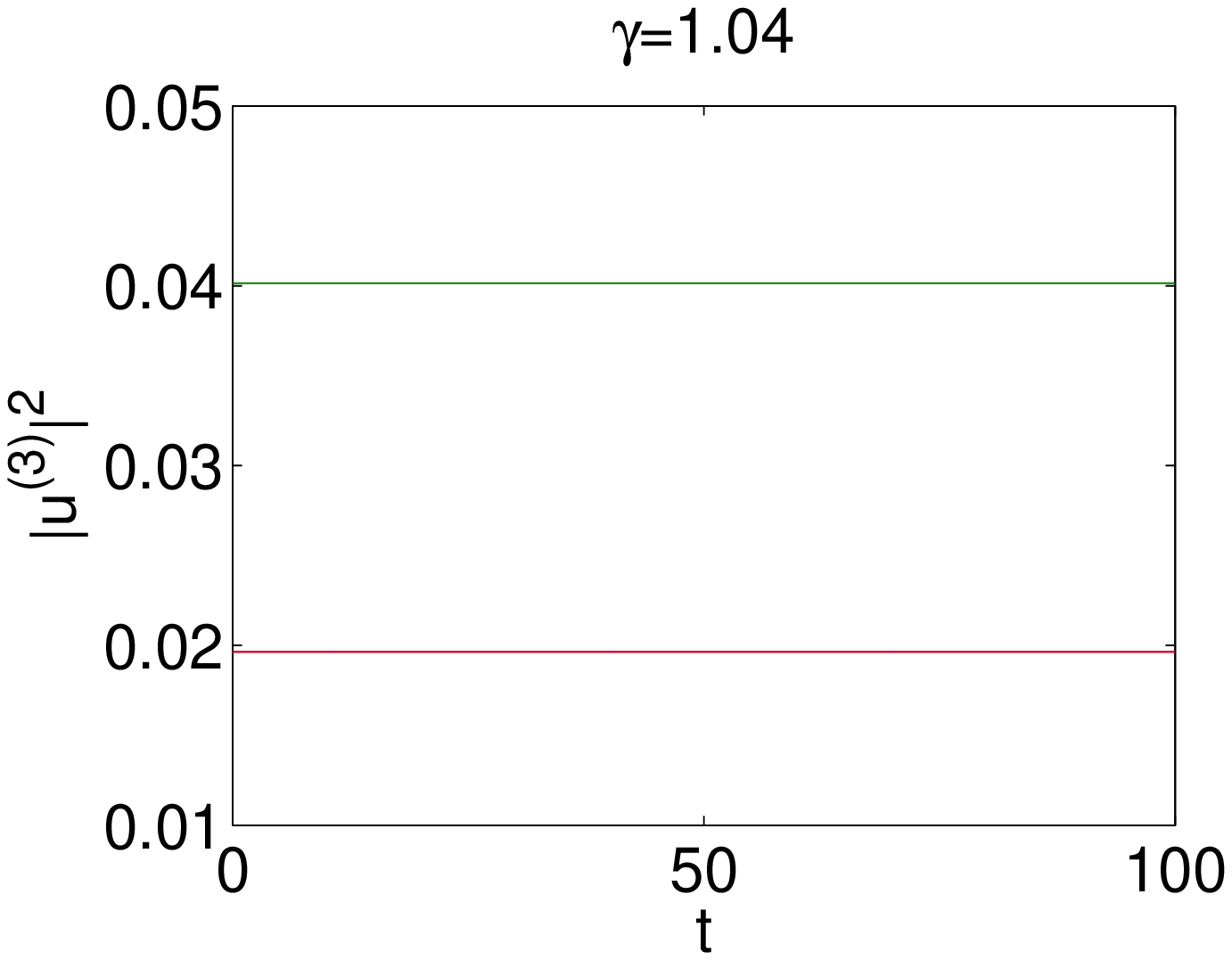}}
\scalebox{0.28}{\includegraphics{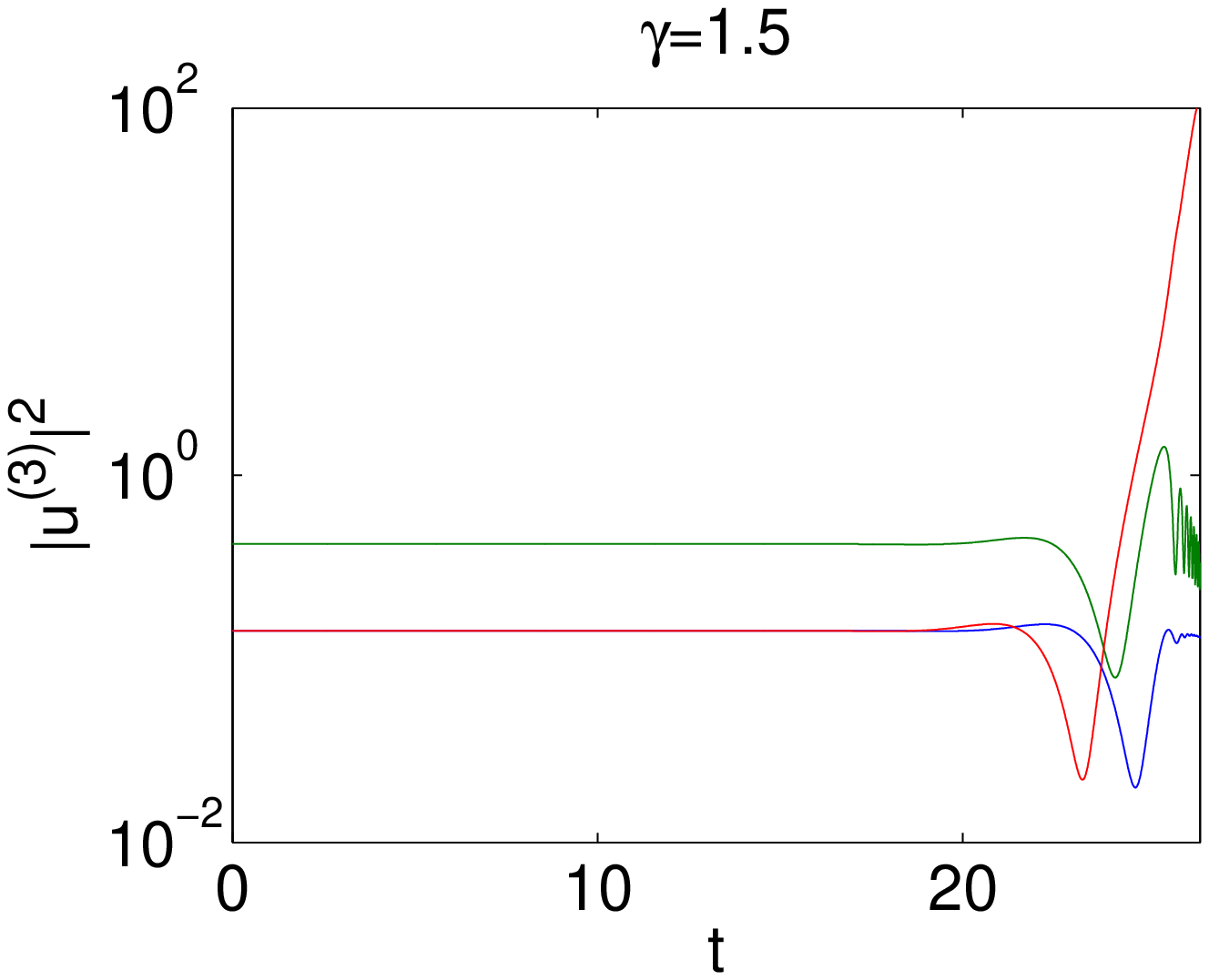}}
\caption{The dynamical evolution for the case of
 $u^{(3)}$ and for $E=k=1$; $\gamma=1.04$ is used in the
left panel and $\gamma=1.5$ in the right one.}
\label{trimer_u3}
\end{figure}

The evolution of the three distinct branches of solutions, namely
the chiefly unstable one $u^{(1)}$, the chiefly stable one $u^{(2)}$
and finally of the one persisting past the linearly unstable limit
$u^{(3)}$ is shown, respectively, in Figs. \ref{trimer_u1},
\ref{trimer_u2} and \ref{trimer_u3}. It can be
seen that in accordance with the predictions of our linear
stability analysis the first two branches are stable
or unstable in their corresponding regimes, while
past the point of existence of these branches ($\gamma=1.043$)
their evolution gives rise to asymmetric dynamics favoring
the growth of the power in a single site (or in some
cases even in two sites; see e.g. the bottom panels of
Figs. \ref{trimer_u1} and \ref{trimer_u2}). On the other hand,
for the branch emerging at $\gamma=1$ and persisting past
the linear instability limit, we indeed find it to be stable
for $\gamma<1.13$ and unstable thereafter again leading to
an asymmetric distribution of the power.

\section{Quadrimer}

Finally, we briefly turn to the case of the
quadrimer. Here the equations are:
\begin{eqnarray}
i\dot{u}_1&=&-ku_2-|u_1|^2u_1-i\gamma u_1
\nonumber
\\
i\dot{u}_2&=&-k(u_1+u_3)-|u_2|^2u_2-i\gamma u_2
\nonumber
\\
i\dot{u}_3&=&-k(u_2+u_4)-|u_3|^2u_3+i\gamma u_3
\nonumber
\\
i\dot{u}_4&=&-ku_3-|u_4|^2u_4+i\gamma u_4
\label{quad1}
\end{eqnarray}
Notice here that we only consider the case where the first
two sites have the same loss and the latter two the same gain.
This is by no means necessary and the gain-loss profile can
be generalized to involve two-parameters (e.g. $\pm \gamma_1$
and
$\pm \gamma_2$ distinct between the different corresponding sites i.e.,
the first and fourth ones, as well as the second and third
ones). We do not consider this latter case here, due to
its more complicated algebraic structure that does not permit
the direct analytical results
given below. More specifically, in our considered
special case, the stationary equations read:
\begin{eqnarray}
E a&=&k b+|a|^2a+i\gamma a
\nonumber
\\
E b&=&k (a+c)+|b|^2b+i\gamma b
\nonumber
\\
E c&=&k (b+d)+|c|^2c-i\gamma c
\nonumber
\\
E d&=&k c+|d|^2d-i\gamma d
\label{quad2}
\end{eqnarray}

The polar representation of the form $a=Ae^{i\phi_a}, b=Be^{i\phi_b},
c=Ce^{i\phi_c}, d=De^{i\phi_d}$ now allows the following reduced
algebraic equations:
\begin{eqnarray}
A^2+B^2=C^2+D^2=E
\label{quad3}
\\
A^2 B^4+\gamma^2A^2-k^2B^2=0
\label{quad4}
\\
D^2 C^4+\gamma^2D^2-k^2C^2=0
\label{quad5}
\\
\sin(\phi_b-\phi_a)=-\frac{\gamma A}{k B}
\label{quad6}
\\
\sin(\phi_c-\phi_b)=-\frac{\gamma E}{k B C}=-1
\label{quad7}
\\
\sin(\phi_d-\phi_c)=-\frac{\gamma D}{k C}.
\label{quad8}
\end{eqnarray}
Notice that in this case not only do we have the customary
phase profile, but in fact one of the phase differences
becomes locked to $\pi/2$ due to the presence of the
gain-loss pattern.

Upon reducing the algebraic equations, we obtain
\begin{eqnarray}
(E-B^2)B^4+\gamma^2(E-B^2)-k^2B^2=0 \label{quadrimer_main1}\\
(E-C^2)C^4+\gamma^2(E-C^2)-k^2C^2=0 \label{quadrimer_main2}\\
\gamma E=kBC \label{quadrimer_main3}
\end{eqnarray}
This leads to the important conclusion that for this gain-loss
profile in the case of the quadrimer,
differently than in the cases of the dimer and trimer,
one of the parameters $E,\ k,\ \gamma$ is determined by the other two;
i.e., not all three of these parameters can be picked independently
in order to give rise to a solution of the quadrimer.

\begin{figure}[htp]
\scalebox{0.28}{\includegraphics{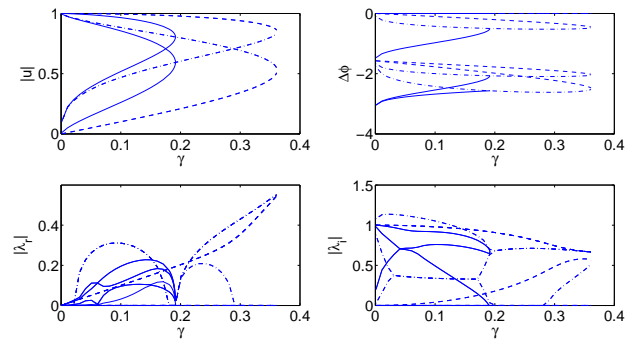}}
\caption{Three branches of solutions
for the trimer problem with parameters $E=1$ and
$\phi_a$ normalized to $0$: the solid lines denote the asymmetric
branch $u^{(1)}$, while the dashed and dash-dotted  denote the symmetric
branches  $u^{(2)}$ and  $u^{(3)}$, respectively.}
\label{quadrimer}
\end{figure}

We hereby set $E=1$, and increase $\gamma$ from $0$ as before, then $k$ can be
obtained self-consistently from the above equations. Therefore, once $E$ and
$\gamma$ are fixed, the solutions of the quadrimer problem are
fully determined. We now present three branches of solutions that arise
in this setting, as we increase $\gamma$. These are shown in
the panels of Fig.~\ref{quadrimer}. There are two classes
of solutions here. The solid curve $u^{(1)}$ corresponds to a fully asymmetric branch with $A$, $B$,
$C$, $D$ distinct, something that is unique (among the settings considered
herein) to the quadrimer. Furthermore, this always unstable branch
does not respect the Hamiltonian eigenalue symmetry i.e., that if
$\lambda$ is an eigenvalue, so are $-\lambda$, $\lambda^{\star}$ and
$-\lambda^{\star}$.
On the other hand, the dashed curve of the
branch $u^{(2)}$
and the dash-dotted curve of $u^{(3)}$ correspond to symmetric
branches with amplitudes $A=D$ and $B=C$. Among the two symmetric branches $u^{(2)}$ and $u^{(3)}$ that collide and disappear together in a saddle-center bifurcation
at $\gamma=0.362$,
we can observe that the former one between them has a real and two imaginary
pairs of eigenvalues being always unstable, while the latter starts
out stable, but the collision of two of its imaginary
pairs will render it unstable past the critical point of $\gamma=0.023$.
Interestingly the asymmetric branch $u^{(1)}$ and the symmetric branch $u^{(3)}$
appear to collide in a subcritical
pitchfork bifurcation that imparts the instability
of the asymmetric branch to the symmetric one for $\gamma>0.193$.

As an aside, we should also note here that in its
linear dynamics (examined e.g. experimentally in
\cite{kip}) the PT-symmetric quadrimer has an interesting difference
from the dimer and trimer. In particular, the 4 linear eigenvalues
of the system are:
\begin{eqnarray}
\lambda_{1,2}=\pm \sqrt{-\gamma^2 + \frac{k}{2} \left(3 k -
\sqrt{-16 \gamma^2 + 5 k^2}\right)}
\label{pteq13}
\\
\lambda_{3,4}=\pm \sqrt{-\gamma^2 + \frac{k}{2} \left(3 k +
\sqrt{-16 \gamma^2 + 5 k^2}\right)}.
\label{pteq14}
\end{eqnarray}
The fundamental difference of this case from the others considered
above is that these eigenvalues do not become imaginary
by crossing through 0. Instead, they become {\it genuinely complex},
through their collision which occurs for  $\gamma=\sqrt{5} k/4$,
a critical point which is lower than that of the trimer. This could
be an experimentally observable signature of the difference between
the near linear dynamics of the quadrimer in comparison e.g. to the
trimer.

\begin{figure}[htp]
\scalebox{0.28}{\includegraphics{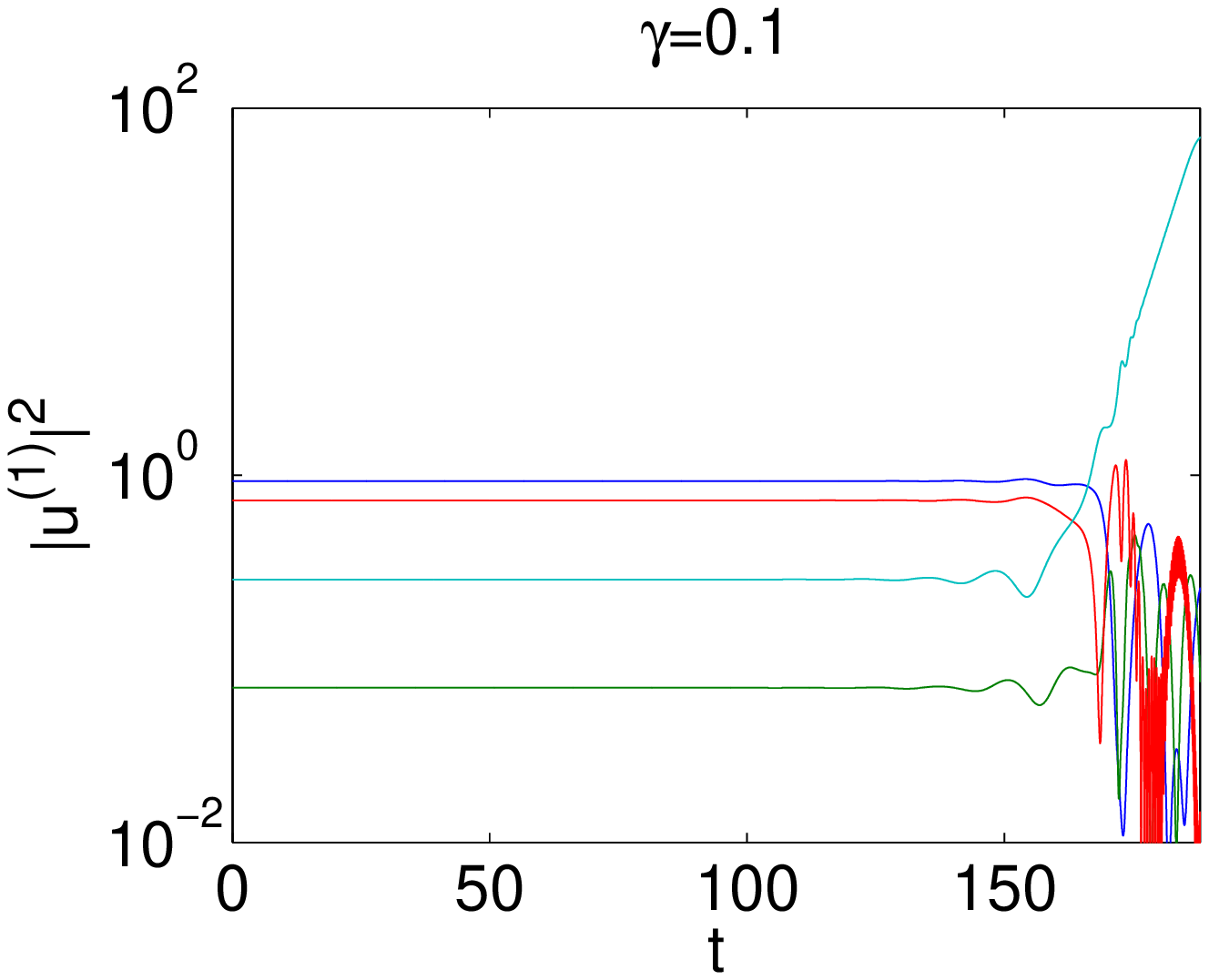}}
\scalebox{0.28}{\includegraphics{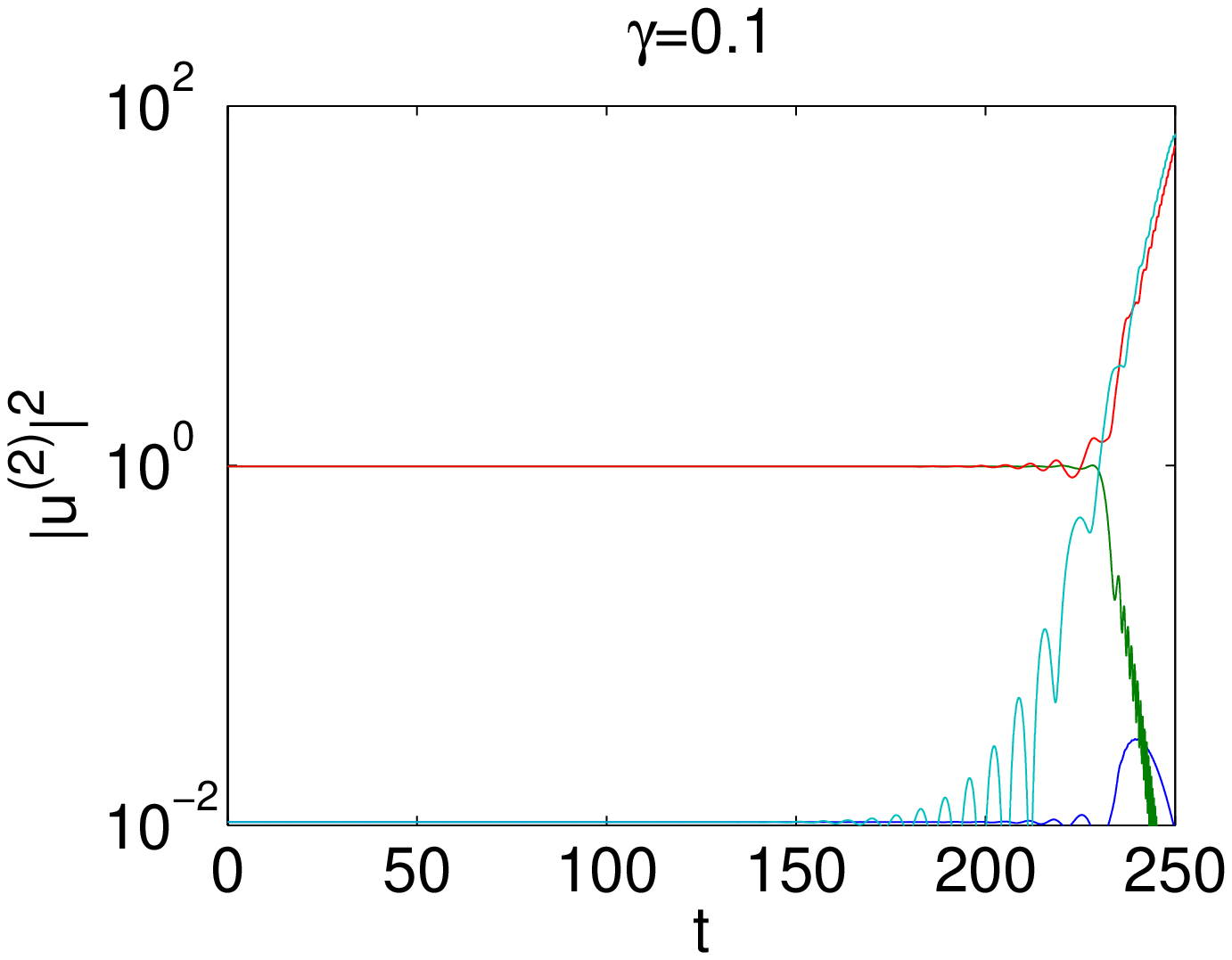}}
\scalebox{0.28}{\includegraphics{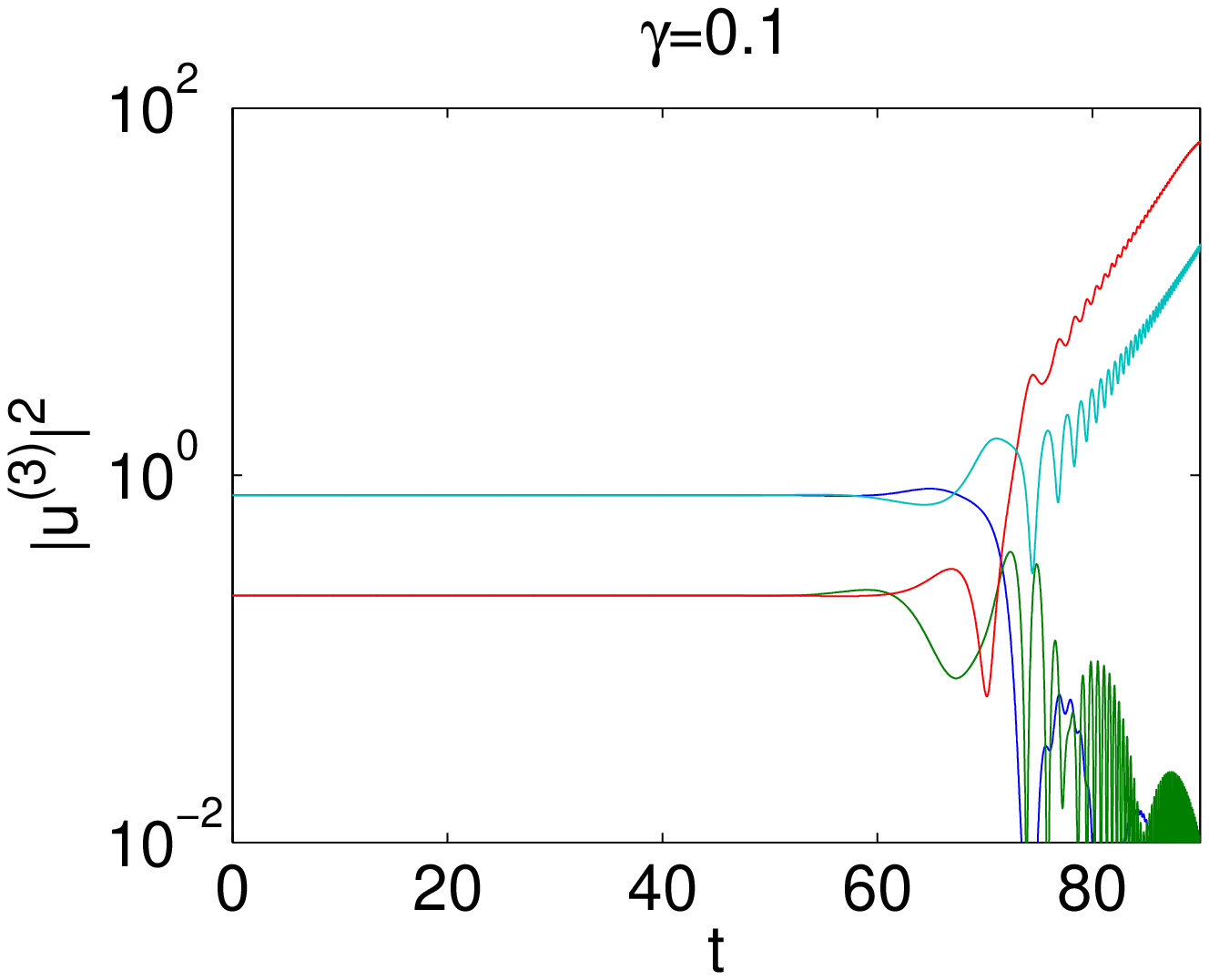}}
\caption{The profile of the dynamical evolution of the three
different branches (top left for $u^{(1)}$, top right for $u^{(1)}$
and bottom panel for $u^{(3)}$)  of a quadrimer in the case of $E=1$
and $\gamma=0.1$.}
\label{quadrimer_RK4}
\end{figure}

The dynamics of these different branches was also considered in Fig.~\ref{quadrimer_RK4}.
In this case, it can be clearly observed that all three branches
tend towards an asymmetric distribution of the power. This favors the
two sites (third and fourth) with the gain, although some case 
examples can be found (see e.g. the top left panel of Fig.~\ref{quadrimer_RK4}
for the asymmetric branch), where only one of the two gain sites is favored
by the mass evolution.

\section{Conclusions}

In the present work, we considered the existence, stability and
dynamics of PT-symmetric oligomers i.e., configurations with few
sites. Similarly to the recent works of \cite{kot1,sukh1} and also
the experimental investigation of \cite{kip}, we have started
our considerations by a complete characterization of the dimer
case, where the two obtained branches of solutions terminate
at the critical point of the linear case. However, we illustrated
that the trimer and quadrimer feature a number of fundamental
differences in comparison with this dimer behavior. In particular,
the trimer features branches which exist past the linear critical
point (although unstable). On the other hand, the quadrimer has
even richer features: in particular, it possesses asymmetric
solutions whose spectrum only has symmetry around the x-axis
(and not the four-fold symmetry of the $\gamma=0$ Hamiltonian problem).
The bifurcation structure is also richer in the latter problem
featuring symmetry-breaking pitchfork bifurcations.
Another notable feature is that solutions do not exist
for arbitrary combinations of coupling, gain/loss parameter
and propagation constant; instead, these parameters appear to
be inter-connected (at least in the case of a single gain-loss
parameter considered herein).
Finally, even the linear problem presents interesting variations
in this case, featuring the breaking of the real nature of the
eigenvalues through two colliding pairs that lead to a quartet
occuring for smaller gain/loss parameter values than in the trimer
case.

This investigation may be a first step towards obtaining
a deeper analytical understanding of the features of
PT-symmetric lattices. In such settings
it would be relevant to obtain general conclusions both for
the linear dynamics (and how it depends on the gain/loss profile
parameters) as well as more importantly for the nonlinear modes,
including the solitary waves that may arise. Understanding such
modes and the comparison of their properties to the continuum ones, as well
as to the discrete ones in the absence of the gain/loss would
be important directions for future study.

\acknowledgments PGK gratefully acknowledges the support of
NSF grants: NSF-DMS-0806762, NSF-CMMI-1000337 and of the Alexander von
Humboldt Foundation as well as of the Alexander S. Onassis
Public Benefit Foundation. He also acknowledges a number of
useful discussions with Prof. T. Kottos.


\begin{thebibliography}{99}

\bibitem{reviews} S. Aubry,  Physica D {\bf 103},
201, (1997); S. Flach and C.R. Willis,  Phys. Rep.
{\bf 295} 181 (1998);
D. Hennig and G. Tsironis,
Phys. Rep. {\bf 307}, 333 (1999);
 P.G. Kevrekidis, K.O. Rasmussen, and A.R.
Bishop,  Int. J. Mod. Phys. B {\bf 15}, 2833 (2001).
A. Gorbach and S. Flach,
Phys. Rep. {\bf 467}, 1 (2008).

\bibitem{reviews1} D. N. Christodoulides, F. Lederer and Y. Silberberg,
Nature \textbf{424}, 817 (2003); Yu. S. Kivshar and G. P. Agrawal,
\textit{Optical Solitons: From Fibers to Photonic Crystals},
Academic Press (San Diego,
2003).


\bibitem{reviews2} P.G. Kevrekidis and D.J. Frantzeskakis,
Mod. Phys. Lett. B {\bf 18}, 173 (2004).
V.V. Konotop and V.A. Brazhnyi, Mod. Phys. Lett. B {\bf 18} 627, (2004);
O. Morsch and M. Oberthaler, Rev. Mod. Phys. {\bf 78}, 179 (2006). 

\bibitem{reviews3} M. Peyrard, Nonlinearity {\bf 17}, R1 (2004).

\bibitem{pgk_rev} P.G. Kevrekidis, arXiv:1009.3178
(IMA J. Appl. Math., in press).

\bibitem{maniadis} P. Maniadis and S. Flach,
Europhys. Lett. {\bf 74}, 452 (2006).

\bibitem{lars} J. Cuevas, L.Q. English, P.G. Kevrekidis and M.
Anderson, Phys. Rev. Lett. {\bf 102}, 224101 (2009).

\bibitem{kip} C.E. R{\"u}ter, K.G. Makris, R. El-Ganainy,
D.N. Christodoulides, M. Segev, D. Kip,
Nature Phys. {\bf 6}, 192 (2010).

\bibitem{bend} C.M. Bender and S. Boettcher,
Phys. Rev. Lett. {\bf 80}, 5243 (1998); C.M. Bender, S. Boettcher
and P.N. Meisinger, J. Math. Phys. {\bf 40}, 2201 (1999);
C.M. Bender, Rep. Prog. Phys. {\bf 70}, 947 (2007).

\bibitem{christo1} Z.H. Musslimani, K.G. Makris, R. El-Ganainy
and D.N. Christodoulides, Phys. Rev. Lett. {\bf 100}, 030402 (2008);
K.G. Makris, R. El-Ganainy, D.N. Christodoulides and Z.H. Musslimani,
Phys. Rev. A {\bf 81}, 063807 (2010).

\bibitem{kot1} H. Ramezani, T. Kottos, R. El-Ganainy and D.N.
Christodoulides, Phys. Rev. A {\bf 82}, 043803 (2010).

\bibitem{sukh1} A.A. Sukhorukov, Z. Xu and Yu.S. Kivshar,
Phys. Rev. A {\bf 82}, 043818 (2010).

\bibitem{kot2} M.C. Zheng, D.N. Christodoulides, R. Fleischmann
and T. Kottos, Phys. Rev. A {\bf 82}, 010103(R) (2010).

\bibitem{grae1} E.M. Graefe, H.J. Korsch and A.E. Niederle,
Phys. Rev. Lett. {\bf 101}, 150408 (2008).

\bibitem{grae2} E.M. Graefe, H.J. Korsch and A.E. Niederle,
Phys. Rev. A {\bf 82}, 013629 (2010).

\end{thebibliography}
\end{document}